\documentclass{article}

\def\xxinput#1{\input#1}

\xxinput{vsolj02.sty}

\usepackage[dvipdfmx]{graphicx}
\usepackage[comma,colon]{natbib}

\usepackage[OT2,T1]{fontenc}

\def\cite{\citealt}
\setcitestyle{aysep={}}

\newcounter{author}
\def\altaffilmark#1{$^{#1}$}
\def\altaffiltext#1{$^{#1}$\,}

\def\authorcount#1#2{{\refstepcounter{author}\label{#1}
                     \altaffiltext{\ref{#1}}{#2}}}

\begin{document}

\begin{center}

\title{On the orbital period of the dwarf nova CW Mon}

\author{
        Taichi~Kato\altaffilmark{\ref{affil:Kyoto}},
        Franz-Josef~Hambsch\altaffilmark{\ref{affil:GEOS}}$^,$\altaffilmark{\ref{affil:BAV}}$^,$\altaffilmark{\ref{affil:Hambsch}}}
\authorcount{affil:Kyoto}{
     Department of Astronomy, Kyoto University, Sakyo-ku,
     Kyoto 606-8502, Japan}
\email{tkato@kusastro.kyoto-u.ac.jp}

\authorcount{affil:GEOS}{
     Groupe Europ\'een d'Observations Stellaires (GEOS),
     23 Parc de Levesville, 28300 Bailleau l'Ev\^eque, France}

\authorcount{affil:BAV}{
     Bundesdeutsche Arbeitsgemeinschaft f\"ur Ver\"anderliche Sterne
     (BAV), Munsterdamm 90, 12169 Berlin, Germany}

\authorcount{affil:Hambsch}{
     Vereniging Voor Sterrenkunde (VVS), Oostmeers 122 C,
     8000 Brugge, Belgium}

\end{center}

\begin{abstract}
\xxinput{abst.inc}
\end{abstract}

   CW Mon is a relatively bright and nearby ($\sim$330~pc,
\cite{GaiaEDR3}) SS Cyg-type dwarf nova.  The orbital period
of this object was assumed to be 0.1766~d in \citet{kat03cwmon}
based on a preliminary period of 0.176~d and the possible
detection of grazing eclipses by \citet{szk86cwmonxleoippegafcamIR}.
This period was listed in \citet{RKCat} and in AAVSO VSX
\citep{wat06VSX},\footnote{
   $<$https://www.aavso.org/vsx/index.php?view=detail.top\&oid=18951$>$.
} and has widely been referenced
(e.g., \cite{war04CVoscillation,pre08CVHalpha,
har16CVsecabundance,hau17cwmon,deb18CVgaia}).

\begin{figure*}
\begin{center}
\includegraphics[width=14cm]{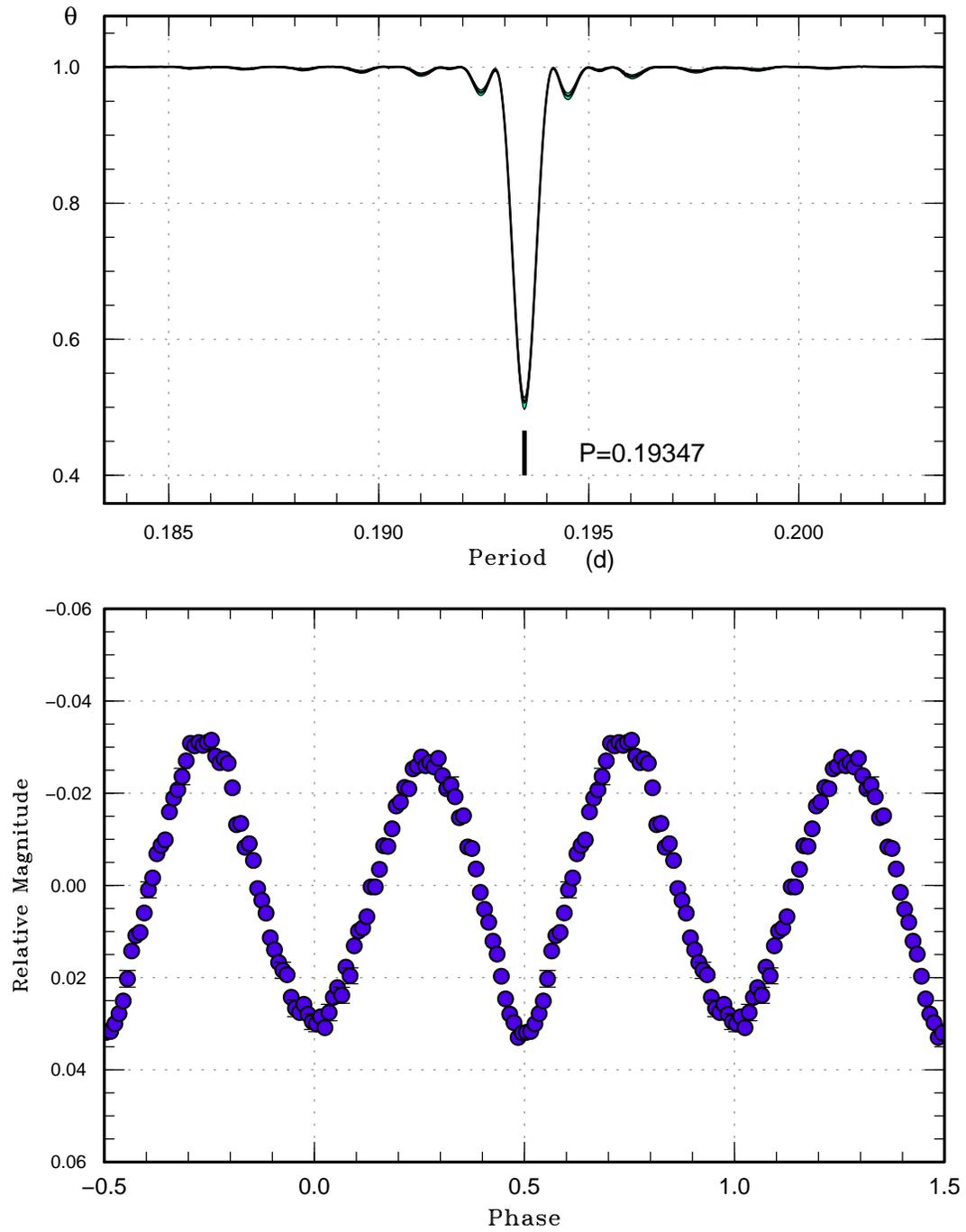}
\caption{
   Period analysis of the TESS data in quiescence.
   (Upper): PDM analysis.  The bootstrap result using
   randomly contain 50\% of observations is shown as
   a form of 90\% confidence intervals in the resultant 
   $\theta$ statistics.
   (Lower): Phase plot.  The zero phase (likely inferior conjunction
   of the secondary) is BJD 2459214.651.
}
\label{fig:porb}
\end{center}
\end{figure*}

\begin{figure*}
\begin{center}
\includegraphics[width=14cm]{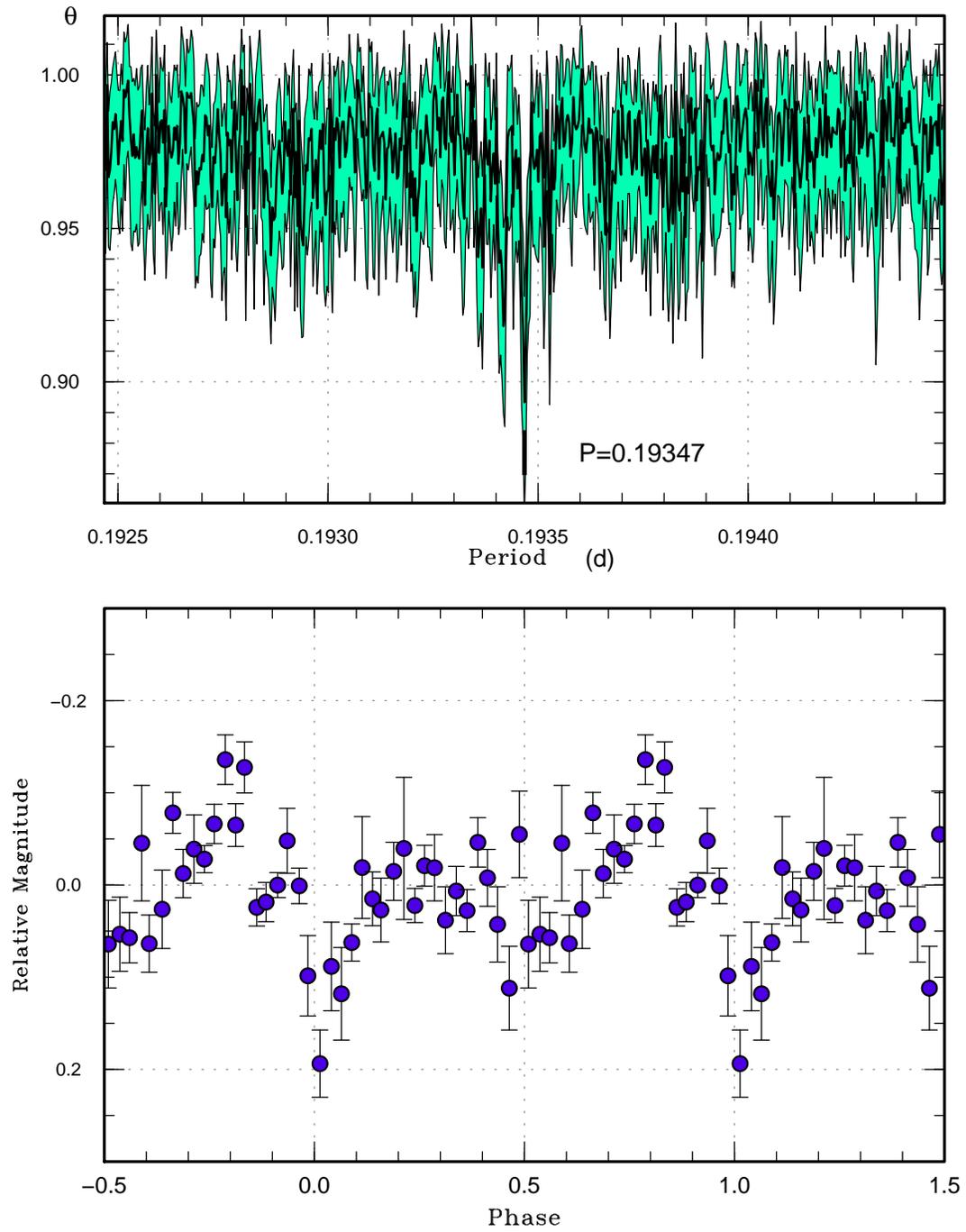}
\caption{
   Period analysis of the ZTF data in quiescence.
   (Upper): PDM analysis.
   (Lower): Phase plot.  The zero phase is the same as in
   figure \ref{fig:porb}.
}
\label{fig:ztfporb}
\end{center}
\end{figure*}

   One of the authors (TK) found a period different from this using
Transiting Exoplanet Survey Satellite (TESS) observations
\citep{ric15TESS}\footnote{
  $<$https://tess.mit.edu/observations/$>$.
  The full light-curve
  is available at the Mikulski Archive for Space Telescope
  (MAST, $<$http://archive.stsci.edu/$>$).
} in quiescence between 2020 December 18 and
2021 January 13.  These observations started $\sim$10~d
after fading of the 2020 November--December outburst.
No further outburst was recorded by 2021 May 4
[The data sources of the long-term behavior were
the All-Sky Automated Survey for Supernovae
(ASAS-SN: \cite{ASASSN}), the Asteroid Terrestrial-impact
Last Alert System (ATLAS: \cite{ATLAS}) forced photometry
\citep{shi21ALTASforced},
the Zwicky Transient Facility (ZTF: \cite{ZTF})\footnote{
   The ZTF data can be obtained from IRSA
$<$https://irsa.ipac.caltech.edu/Missions/ztf.html$>$
using the interface
$<$https://irsa.ipac.caltech.edu/docs/program\_interface/ztf\_api.html$>$
or using a wrapper of the above IRSA API
$<$https://github.com/MickaelRigault/ztfquery$>$.
} and observations reported to VSOLJ and VSNET \citep{VSNET}].
The presence of the orbital variation was already visible
to the eyes in the TESS light curve.  After removing the global
trend by locally-weighted polynomial regression
(LOWESS: \cite{LOWESS}), a phase dispersion minimization
(PDM, \cite{PDM}) analysis yielded an orbital period
(figure \ref{fig:porb}) having a period of 0.193467(4)~d
with its error determined by the methods of \citet{fer89error} 
and \citet{Pdot2}.  The zero orbital phase (BJD 2459214.651)
was chosen as the shallower minimum of the ellipsoidal
variations, which likely corresponds to the inferior conjunction
of the secondary reflecting the effect of
gravitational darkening.  The amplitudes of the ellipsoidal
variations were small in the TESS data due to contamination
of nearby stars.  This period could not be detected
in the ATLAS forced photometry data (2145 measurements).
A PDM analysis of the ZTF data
(quiescent parts only, after removing the trends by LOWESS)
detected a period of 0.193468(1)~d (figure \ref{fig:ztfporb})
(due to the large scatter in the individual data, we only show
a phase-averaged light curve of the combined $g$+$r$ data).
This light curve appears to show an orbital hump.

\begin{figure*}
\begin{center}
\includegraphics[width=14cm]{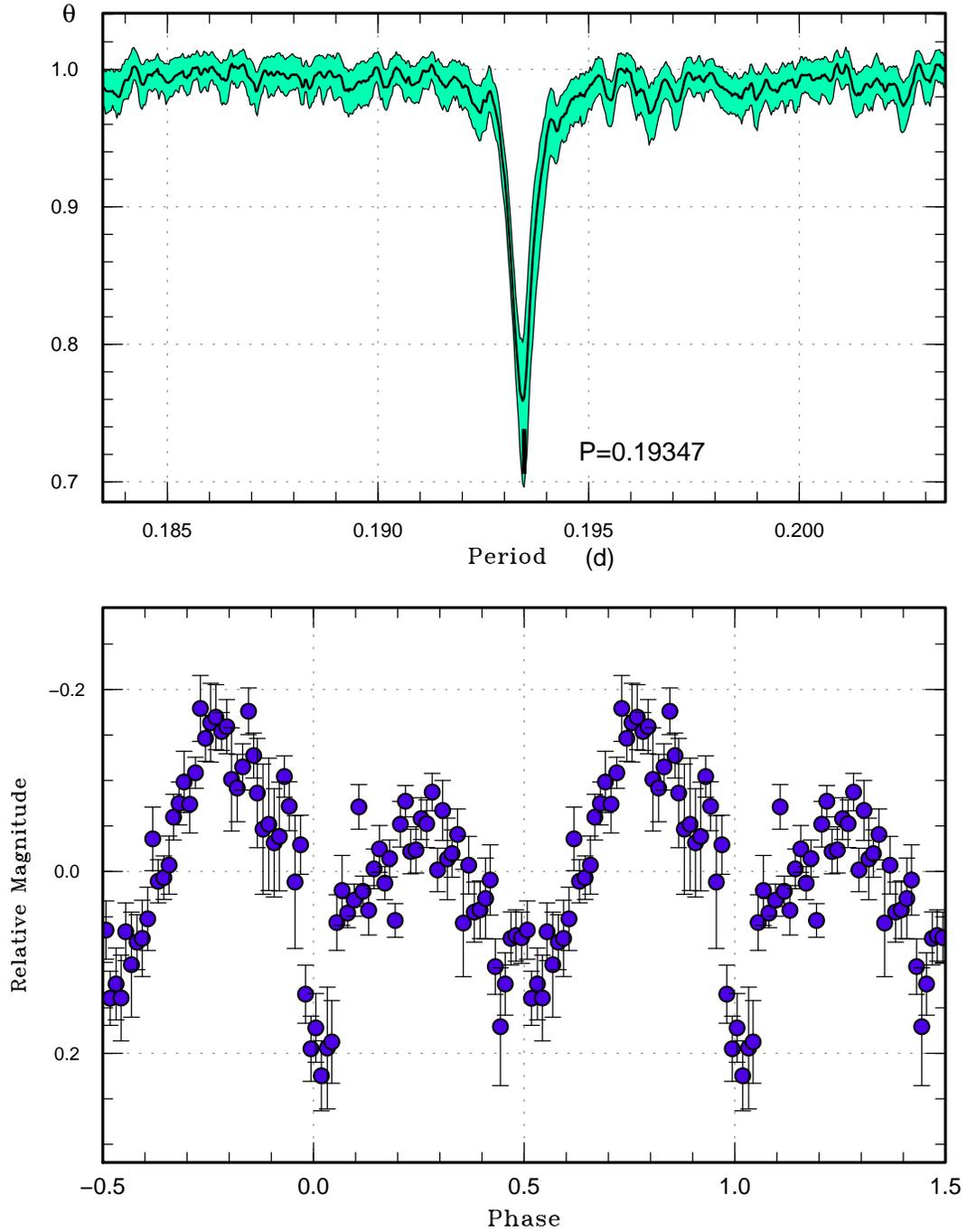}
\caption{
   Period analysis of the data before and after
   the 2016 October outburst (BJD 2457669--2457671 and
   BJD 2457683--2457711).
   (Upper): PDM analysis.  The vertical tick is given at
   the orbital period derived from the entire data discussed
   in this paper.
   (Lower): Phase plot.  The zero phase is the same as in
   figure \ref{fig:porb}.  An eclipse and an orbital hump are
   apparently seen.
}
\label{fig:porb2016}
\end{center}
\end{figure*}

   The same period was detected from the data obtained
during and around the 2016 October outburst (observer: FJH).
The phase-averaged light curve (figure \ref{fig:porb2016})
in quiescence before and after the outburst indicates
the presence of an eclipse and an orbital hump.
This finding apparently reinforces the presence of grazing eclipses
reported during the 2002 outburst \citep{kat03cwmon}.
Using all the data (TESS, ZTF and the 2016 VSNET campaign)
between 2002 and 2022, we obtained a refined ephemeris of:
\begin{equation}
\mathrm{Min(BJD)} = 2459214.651 + 0.19346802(4) E.
\end{equation}

\begin{figure*}
\begin{center}
\includegraphics[width=14cm]{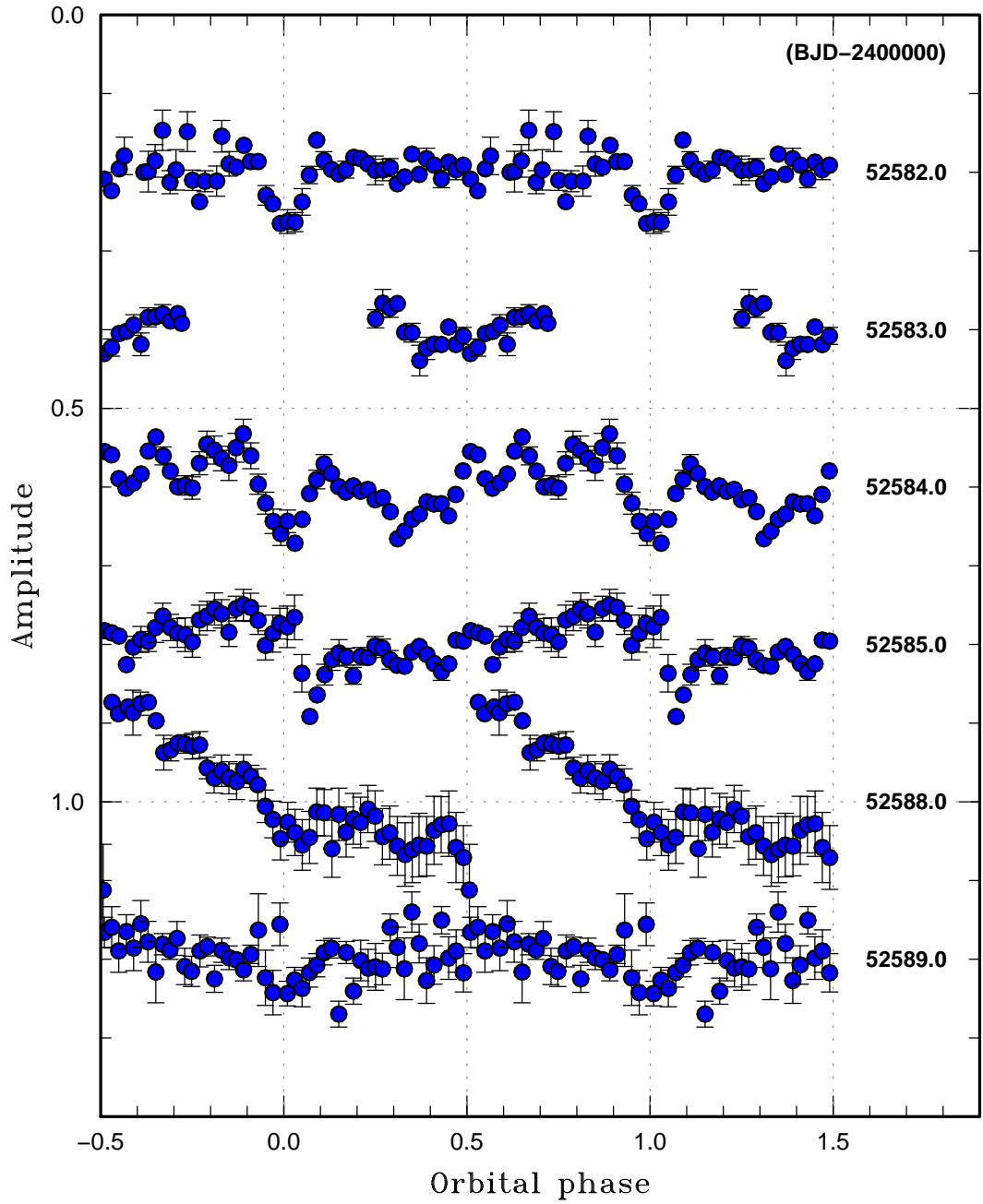}
\caption{
   Nightly orbital profiles during the 2002 outburst using
   the correct orbital period.  A discontinuity in the light
   curve of BJD 2452588.0 was an artificial one caused by
   two different observers giving different trends.
}
\label{fig:prof}
\end{center}
\end{figure*}

\begin{figure*}
\begin{center}
\includegraphics[width=14cm]{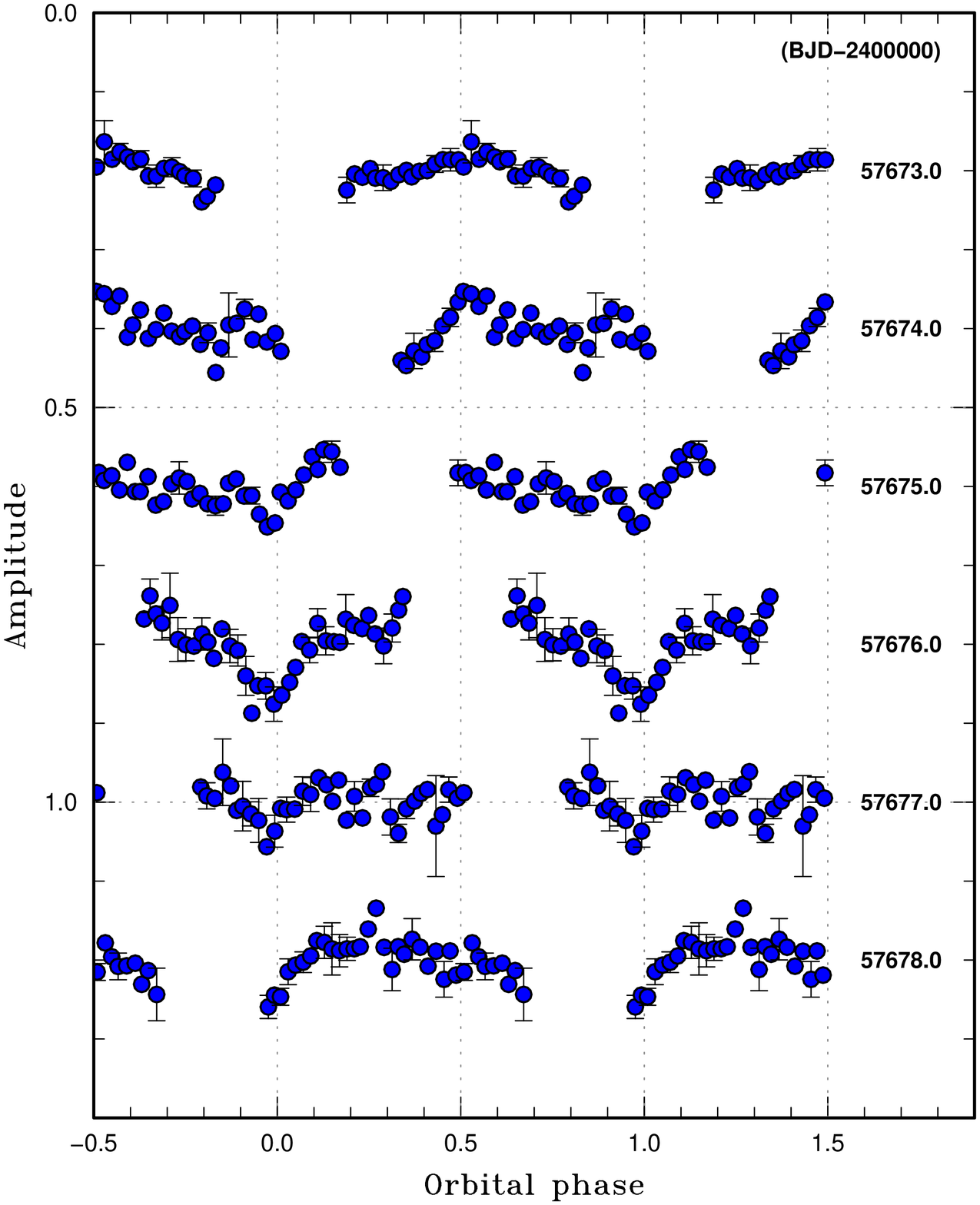}
\caption{
   Nightly orbital profiles during the 2016 outburst.
   Eclipses were visible at least on the third to fifth
   nights.
}
\label{fig:prof2016}
\end{center}
\end{figure*}

   Using this ephemeris, the corrected figure corresponding
to figure 4 in \citet{kat03cwmon} is shown in figure
\ref{fig:prof}.  Although the disappearance of an eclipse on
the second night during the 2002 outburst was likely due to
the usage of an incorrect orbital period, not the change
in the disk radius suggested in \citet{kat03cwmon},
this could not be directly confirmed due to the absence of
the data on the second night around the expected eclipse.
The two periods of 0.1766~d and 0.19346802~d are
in the relation of 2-day alias, which was probably introduced by
a confusion between the two maxima/minima of the ellipsoidal
variations by \citet{szk86cwmonxleoippegafcamIR}.
The same figure for the 2016 outburst is shown in
figure \ref{fig:prof2016}.  Eclipses were also present.
The available data seem to confirm that CW Mon is
a grazing eclipser both in outburst and in quiescence,
at least in quiescence before and after the 2016 outburst.
The eclipse signature in the TESS data was less apparent
than in 2016.  Although there were some observations
during other outbursts in 2005 (Hiroyuki Maehara), 2008
(Seiichiro Kiyota) and 2010 (S. Kiyota),
the results were not as clear as in 2016.
An eclipse-like signal was possibly present
at the expected phase on two nights, but absent in
one night in 2005.  The data were insufficient in 2008
and the eclipse signature was apparently absent in 2010
in one night covering this phase.  These observations
were relatively short and did not cover much of outbursts
(and they did not record quiescence before and after the outburst)
as in 2016, and we feel it difficult to draw a firm conclusion.
It is possible that the eclipse signature is missing
during some outbursts.  The 2002 and 2016 outbursts had
a ``shoulder'', or an embedded precursor \citep{can12ugemLC},
sometimes seen in long outbursts of dwarf novae above
the period gap and the disk was probably large during
such outbursts.
In other outbursts, either the size of the disk may
have not been sufficient or the outer part of the disk
may have not been bright enough to show detectable eclipses.
This interpretation may be tested by observations of
future observations of different types of outbursts.

   Regarding the possible signal of an intermediate polar
\citep{kat03cwmon}, \citet{pre08CVHalpha} did not detect
a coherent pulsation nor a spectroscopic signature of
a magnetized white dwarf.  \citet{war04CVoscillation}
classified the reported 37-min signal as quasi-periodic
oscillations (QPOs).  There has been increasing evidence
that short-term variations are excited around a shoulder
(or an embedded precursor).
For example, quite recently, \citet{sun23hs2325}
reported enhancement of QPO signals near the long outburst
top light curves in HS 2325$+$8205 = NSV 14581 using the TESS data.
These outbursts had a shoulder.  Similar enhancement of short-term
variations (an orbital or a superhump signal) during a shoulder
of a long outburst was also reported in V363 Lyr \citep{kat21v363lyr}
using the Kepler data.  The QPOs observed in CW Mon
might have been related to these phenomena, and may be
related to excitation of short-term variations when the accretion
disk reaches the maximum radius, the tidal truncation radius
as being a possibility.

\section*{Acknowledgements}

This work was supported by JSPS KAKENHI Grant Number 21K03616.

The author is grateful to the TESS, ATLAS, ZTF and ASAS-SN
teams for making their data available to the public.
I am grateful to Naoto Kojiguchi for helping downloading
the TESS data.  I am also grateful to VSOLJ and VSNET
observers who reported the data of CW Mon, particularly
to Hiroyuki Maehara and Seiichiro Kiyota for providing
time-resolved photometric data during past outbursts.

This work has made use of data from the Asteroid Terrestrial-impact
Last Alert System (ATLAS) project. The Asteroid Terrestrial-impact
Last Alert System (ATLAS) project is primarily funded to search for
near earth asteroids through NASA grants NN12AR55G, 80NSSC18K0284,
and 80NSSC18K1575; byproducts of the NEO search include images and
catalogs from the survey area. This work was partially funded by
Kepler/K2 grant J1944/80NSSC19K0112 and HST GO-15889, and STFC
grants ST/T000198/1 and ST/S006109/1. The ATLAS science products
have been made possible through the contributions of the University
of Hawaii Institute for Astronomy, the Queen's University Belfast, 
the Space Telescope Science Institute, the South African Astronomical
Observatory, and The Millennium Institute of Astrophysics (MAS), Chile.

Based on observations obtained with the Samuel Oschin 48-inch
Telescope at the Palomar Observatory as part of
the Zwicky Transient Facility project. ZTF is supported by
the National Science Foundation under Grant No. AST-1440341
and a collaboration including Caltech, IPAC, 
the Weizmann Institute for Science, the Oskar Klein Center
at Stockholm University, the University of Maryland,
the University of Washington, Deutsches Elektronen-Synchrotron
and Humboldt University, Los Alamos National Laboratories, 
the TANGO Consortium of Taiwan, the University of 
Wisconsin at Milwaukee, and Lawrence Berkeley National Laboratories.
Operations are conducted by COO, IPAC, and UW.

The ztfquery code was funded by the European Research Council
(ERC) under the European Union's Horizon 2020 research and 
innovation programme (grant agreement n$^{\circ}$759194
-- USNAC, PI: Rigault).

\section*{List of objects in this paper}
\xxinput{objlist.inc}

\section*{List of objects in this paper}
\xxinput{objlist.inc}

\section*{References}

We provide two forms of the references section (for ADS
and as published) so that the references can be easily
incorporated into ADS.

\newcommand{\noop}[1]{}\newcommand{\hyphalt}{-}

\renewcommand\refname{\textbf{References (for ADS)}}

\xxinput{cwmonaph.bbl}

\renewcommand\refname{\textbf{References (as published)}}

\xxinput{cwmon.bbl.vsolj}

\end{document}